\documentclass[conference]{IEEEtran}
\IEEEoverridecommandlockouts

\usepackage{cite}
\ifCLASSINFOpdf
   \usepackage[pdftex]{graphicx}
\else
  \usepackage[dvips]{graphicx}
\fi
\usepackage[cmex10]{amsmath}
\usepackage{url,color,cite,booktabs}

\def\beq{\begin{equation}}
\def\eeq{\end{equation}}
\def\beqa{\begin{eqnarray}}
\def\eeqa{\end{eqnarray}}

\def\MT{{\ensuremath{\mathit{MT}}}}
\def\IP{{\ensuremath{\mathit{IP}}}}
\def\TP{{\ensuremath{\mathit{TP}}}}

\begin{document}

\title{Let Us Dance Just a Little Bit More --- On the Information
  Capacity of the Human Motor System}

\author{\IEEEauthorblockN{Teemu Roos}
\IEEEauthorblockA{Helsinki Inst. for Inform. Tech. HIIT\\
University of Helsinki\\ 
Helsinki, Finland\\
Email: \url{{firstname.lastname}@hiit.fi}}
\and
\IEEEauthorblockN{Antti Oulasvirta}
\IEEEauthorblockA{Computer Graphics Department\\
Max Planck Institute for Informatics\\
Saarbr\"ucken, Germany\\
Email: \url{oantti@mpi-inf.mpg.de}}
\and
\IEEEauthorblockN{Laura Lepp\"anen and Arttu Modig}
\IEEEauthorblockA{Helsinki Inst. for Inform. Tech. HIIT\\
University of Helsinki\\ 
Helsinki, Finland\\
Email: \url{{firstname.lastname}@hiit.fi}}}

\maketitle

\begin{abstract}
  Fitts’ law is a fundamental tool in measuring the capacity of the human motor system. However, it is, by definition, limited to aimed movements toward spatially expanded targets. We revisit its information-theoretic basis with the goal of generalizing it into unconstrained trained movement such as dance and sports. The proposed new measure is based on a subject’s ability to accurately reproduce a complex movement pattern. We demonstrate our framework using motion-capture data from professional dance performances. 
\end{abstract}

{\keywords Fitts' law, information capacity, 
human motor system, human-computer interaction}

\IEEEpeerreviewmaketitle

\section{Introduction}

The purpose of the human motor system is to transform electro-chemical
signals in the nervous system into physical movement. 
The dominant paradigm for studying the information capacity of the
human motor system is based on the pioneering work by Paul Fitts in
the 1950s~\cite{fit:54,fit:pet:64,wel:68}. Its primary application is
the analysis of user interfaces in human-computer 
interaction~\cite{mac:92,sou:mac:04,zha:02}; it was, for instance, one of
the main drivers in the development and adoption of the computer 
mouse~\cite{atk:07}. 

Fitts was interested in \emph{aimed movements}; i.e., movement where a
pointer (finger, eye fixation, arm, mouse cursor etc.) is moved on top
of a spatially expanded target. A common example is moving mouse cursor
on top of a button on a computer display. Fitts' law describes the
observation that the relationship between movement time $\MT$ and
spatial characteristics of the required movement is
 characterized as: \beq
\MT=a+b\log_2 \left(1+{D \over W}\right),
\label{eq:fitts}
\eeq where $D$ is the distance from the starting point to the center
of the target and $W$ is the width of the target; $a$ and $b$ are
empirical parameters determined by the task, the pointing device, and the
performer.
$\MT$ is typically measured in an empirical procedure involving
rapid responses to spatial targets with experimenter-controlled
characteristics. 

The information-theoretic basis of Fitts' law is centered around the
tradeoff between the speed and accuracy of movements produced by the
motor system. In information-theoretic terms, the \emph{capacity} of
the motor system as a channel of communication is limited by this
tradeoff. Since physical movements are
naturally measured on a continuous (spatial) scale, the measurement of
their information content must involve the determination of their 
accuracy or, as Fitts points out, 
\begin{quote}
\it `` [s]ince measurable aspects of motor responses, such as their
force, direction, and amplitude, are continuous variables, their
information capacity is limited only by the amount of statistical
variability, or noise, that is characteristic of repeated efforts
to produce the same response. ``~\cite{fit:54}
\end{quote}

The information theoretic interpretation of Fitts' law
\cite{fit:54,fit:pet:64,mac:89,wel:68,zha:02}, where information
throughtput is formalized in terms of a standard Gaussian channel, see~\cite{cov:tho:06},
has been immensely popular since it enables the comparison of 
performance across situations with different
characteristics.  The \emph{index of performance} (IP) defines the
information throughput in units of \emph{bits per second}
(bps)%
: \beq \IP=1/b.
\label{eq:ip}
\eeq
IP is argued to be a good metric because, as observed by Fitts, and later many others,
 it tends to stay relatively constant
over a broad range of values of $D$ and $W$~\cite{sou:mac:04,zha:02}, 
providing a natural basis for comparison of pointing devices.   The
mouse, for example, typically reaches ca.\ 4 bps, and joystick ca.\ 2
bps~\cite{sou:mac:04}.

The motivation for the present work is that   
important aspects of the information potential of human motor system are
not covered by the Fitts' law paradigm, and that consequently, the
capacity of human motor system is systematically underestimated --- insofar as
the said paradigm even attempts to estimate the capacity of the whole motor system. 
In fact, Fitts' law and its generalizations are
constrained to \emph{aimed} movements of one (or few) body part(s) in target
conditions that are \emph{prescribed} to a high degree by the experimenter. This has three
important implications. Firstly, the ``information'' that is being measured is
tantamount to the subject's ability to motorically conform to
extrinsic constraints, excluding entirely free movement, i.e., movement
produced irrespective of its absolute position in respect to
perceivable environmental constraints. Such movements are important in
many skilled activities, such as dance and sports. The issue of
underestimation is exacerbated by the empirical paradigm, which
utilizes very simple repetitive movements with simple
trajectories (see~\cite{acc:zha:99}. Secondly, Fitts' law does not account for information in
\emph{simultaneous movement of multiple body parts} (for an exception,
see~\cite{rob:kav:76}). There are 640 muscles, 200-300 joints, and 206
bones in the human body. Obviously we are not able to independently
control each one of them, but some separation is possible; for instance,
the thumb and the index finger can be moved relatively independently
of each other and the three other fingers~\cite{jon:led:06}. Thirdly,
most skilled activities involve compound tasks, with multiple aimed
and other types of movement performed simultaneously and
sequentially. Due to these three limitations, we argue that the Fitts'
law paradigm is not suitable for the study of skilled motor action; 
i.e., precisely the ones that can be expected
to contain the most information!



Extending Fitts' definition, we define
information capacity in terms of the ability to accurately
\emph{reproduce any previously performed movement pattern}.  An infant
is a good example of \emph{low} information capacity. At any moment in time, the infant's
movement can appear complex, but the fact that he or she cannot reproduce it
at will means that the motor system lacks the information capacity to do so.

Our formulation is based on subjects performing arbitrarily complex
un-prescribed movements; Fitts' paradigm, involving only
experimenter-defined pointing tasks, is a special case.  The
formulation can accommodate movement of any duration and composition
and involving contributions of any part of the body.

The rest of the paper is organized as follows.
In Sec.~\ref{sec:measure}, we describe a measure of
shared information between two movement sequences. 
The data and the preprocessing steps are
detailed in Sec.~\ref{sec:data}, and the results of the experiments
are summarized in Sec.~\ref{sec:results}. To conclude, in
Sec.~\ref{sec:conclusion} we discuss potential applications and
outline future work.

\section{Information Measure}
\label{sec:measure}

To quantify the information capacity, it is necessary to separate the
\emph{controlled} aspects of the performed sequence of movements from
the \emph{unintentional} aspects that are unavoidably present in all
motor responses. As discussed above, the strictly defined range of
admissible performances in Fitts' paradigm has a similar function: it
rules out apparently complex, uncontrolled (random) sequences of
movements.  Instead of restricting the allowed movements, we propose
to solve this task by having a sequence \emph{repeated} as exactly as
possible by the same subject.  This makes it possible to obtain an
estimate of the variability of the two patterns, and subtract the
complexity (entropy) due to it from the total complexity of the
repeated performance.  In other words, information is measured by two
aspects of the performance: $i)$ the complexity of a movement pattern,
and $ii)$ the precision with which it can be repeated. To clarify, we 
let the \emph{complexity} of a sequence be given by its entropy\footnote{In the
case of continuous signals, we continue to do so, keeping in mind the
caveats associated with the interpretation of differential 
entropy, see, e.g.~\cite[Chapter 8]{cov:tho:06}.}.

\subsection{The One-Dimensional Case}

For simplicity, we start by treating the one-dimensional case where
each movement sequence is characterized by a single measurement per time
frame. Let $\mathbf{x} = x_{-1},\dots,x_{n}$ denote a sequence where
$x_t$ gives the value of the measured feature at time $t \in
\{-1,\dots,n\}$. We start the sequence from $x_{-1}$ instead of $x_1$
for notational convenience: the first two entries guarantee that an
autoregressive model with a look-back
(lag) of two steps can be fitted to exactly $n$ data points. Similarly,
we denote by $\mathbf{y} = y_{-1},\ldots,y_n$ another movement
sequence of the same length as $\mathbf{x}$.

We assume that both $\mathbf{x}$ and $\mathbf{y}$ follow  a second-order 
autoregressive model 
	\begin{align} x_t &=
\beta_0 + \beta_1 x_{t-1} + \beta_2 x_{t-2} + \epsilon^{(\mathbf{x})}_t,\\
y_t &=
\eta_0 + \eta_1 y_{t-1} + \eta_2 y_{t-2} + \epsilon^{(\mathbf{y})}_t,
\label{eq:lin1}
\end{align}
 where $\beta_0,\beta_1,\beta_2$ and $\eta_0,\eta_1,\eta_2$   are real-valued
parameters to be tuned using least squares.
The second-order model accounts for the
basic physical principle that once the movement vector (including
direction and velocity) is specified, constant movement contains no
information whatsoever.

The errors (or \emph{innovations})
$\epsilon^{(\mathbf{x})}_t$ and $\epsilon^{(\mathbf{y})}_t$ are assumed to be zero
mean Gaussian random variates. Since the two sequences are supposed to be
repetitions of the same movement pattern, we let $\epsilon^{(\mathbf{x})}_t$ and
$\epsilon^{(\mathbf{y})}_t$ be correlated with some correlation coefficient $\rho \in
(-1,1)$. The innovations for different time frames $t\neq t'$ are assumed to be
independent of each other.

Having fitted the parameters to observed sequences, we obtain the residuals
\begin{align}
r^{(\mathbf{x})}_t &= x_t - \hat x_t = x_t - (\hat\beta_0 + \hat\beta_1 x_{t-1}
+ \hat\beta_2 x_{t-2}),\\
r^{(\mathbf{y})}_t &= y_t - \hat y_t = y_t - (\hat\zeta_0 + \hat\zeta_1 y_{t-1}
+ \hat\zeta_2 y_{t-2}),
\end{align}
where $\hat x_t$ and $\hat y_t$ denote the predicted values based on the least
squares estimates $\hat\beta_0$,$\hat\beta_1$,$\hat\beta_2$
and $\hat\eta_0$,$\hat\eta_1$,$\hat\eta_2$, respectively.  

Under the model~(\ref{eq:lin1}), the (differential) entropy of each of the
sequences
can be estimated by plugging the residual variance into the familiar
formula for the Gaussian entropy (see~\cite{cov:tho:06}):
\beq
h(\mathbf{x}) \approx {n\over 2} \log_2 (2\pi e \hat\sigma_{\mathbf{x}}^2),
\quad h(\mathbf{y}) \approx {n\over 2} \log_2 (2\pi e \hat\sigma_{\mathbf{y}}^2),
\label{eq:gaussent}
\eeq
where $\hat\sigma_{\mathbf{y}}^2 = \sum_{t=1}^n (r^{(\mathbf{y})}_t)^2 / n$ is the residual
variance of $\mathbf{x}$ and $\hat\sigma_{\mathbf{y}}^2$ is defined similarly.

The mutual information between the movement sequences, which gives the reduction
in bits in the entropy of one sequence when we are given the other, is now fully
determined by the residuals, and in particular, their correlation
$\rho$:
\beq
I(\mathbf{x}\,;\,\mathbf{y}) = -{n\over 2} \log_2 (1-\rho^2).
\eeq
However, since we do not in general know the true correlation coefficient, we need to
estimate it from the data. Using the empirical correlation coefficient tends to underestimate
the true value, and hence, our mutual information estimate based on it will tend to be
too high. (For instance, even if the true correlation is zero, we will always get an
estimate that is greater than zero.) There
are various ways to compensate for this bias. We adopt an approach similar to
Rissanen's classic two-part approximation to the \emph{stochastic
complexity}~\cite{ris:78}, whereupon the estimated mutual information becomes
\beq 
\hat I(\mathbf{x} \;;\; \mathbf{y}) 
= -{n\over 2} \log_2 \left(
{1-\hat\rho^2} \right)
- {1\over 2} \log_2 n,
\label{eq:muti}
\eeq where the last term will act to overcome
the overestimation of the mutual information due to fitting the correlation parameter to
a finite amount of data (see, e.g.,~\cite{gru:07} for many interesting properties
of the stochastic complexity formula;  those familiar with the concept, may notice that
our penalty term is equal to ${k\over 2}\log_2 n$ with $k=1$ parameters).

The mutual information has
a direct interpretation in terms of the reduction in \emph{bits}
required to encode the sequence $\mathbf{x}$ due to the side
information $\mathbf{y}$ being available.  Since the mutual
information in $\mathbf{x}$ and $\mathbf{y}$ excludes, with high
probability, most of the uncontrolled movements and inaccuracies
which tend not to be repeated when the movement is performed
twice, we
argue that it provides a measure of the controlled information in
$\mathbf{x}$. To achieve high mutual information, a movement has to be
both complex  and accurately controlled  so that it 
can be repeated with high precision.

Finally, we define the observed \emph{throughput} in a sequence
$\mathbf{x}$ conditioned on sequence $\mathbf{y}$ as the estimated mutual
information per second:
\beq
\TP(\mathbf{x} \mid \mathbf{y}) = 
{R \, \hat I(\mathbf{x}\,;\,\mathbf{y}) \over n}
=-{R\over 2} \log_2 \left(
{1-\hat\rho^2} \right)
- {R \over 2n} {\log_2 n} ,
\label{eq:capa}
\eeq
where $R$ denotes the frame rate (frames per second).

\subsection{The Multidimensional Case}

When handling $p$-dimensional sequences, $p>1$, where each time frame $x_t$ is
composed of $p$ measured components (features), $x_t =
(x_t^{(1)},\dots,x_t^{(p)})$, it is not sufficient to simply sum up the information throughput 
in each of the components separately. This would namely
exaggerate the throughput as redundant information that is contained
in more than one component was counted several times.

To reduce the effect of redundant information shared between features, we 
decorrelate the features. To this end, we
first perform principal component analysis (PCA) on movement sequence $\mathbf{x}$. 
We then transform both sequences to obtain two new time series,
$\mathbf{x}'$ and $\mathbf{y}'$ where each frame in each sequence is obtained by a linear
transformation
(the same one for both $\mathbf{x}$ and $\mathbf{y}$) of the corresponding frame in the 
original sequence. Typically most of the variance in
the new sequences is focused on a fraction of the principal components, and we retain
only as many as are required to cover $90$ percent of the variance (of $\mathbf{x}$). 
The newly obtained lower-dimensional sequences are then analysed using the technique 
described above,  and the throughputs are summed up.

\begin{figure*}[!t]
\centering
\setlength{\unitlength}{\textwidth}
\begin{picture}(1,.4)
\put(0,.025){
\includegraphics[height=1.9in]{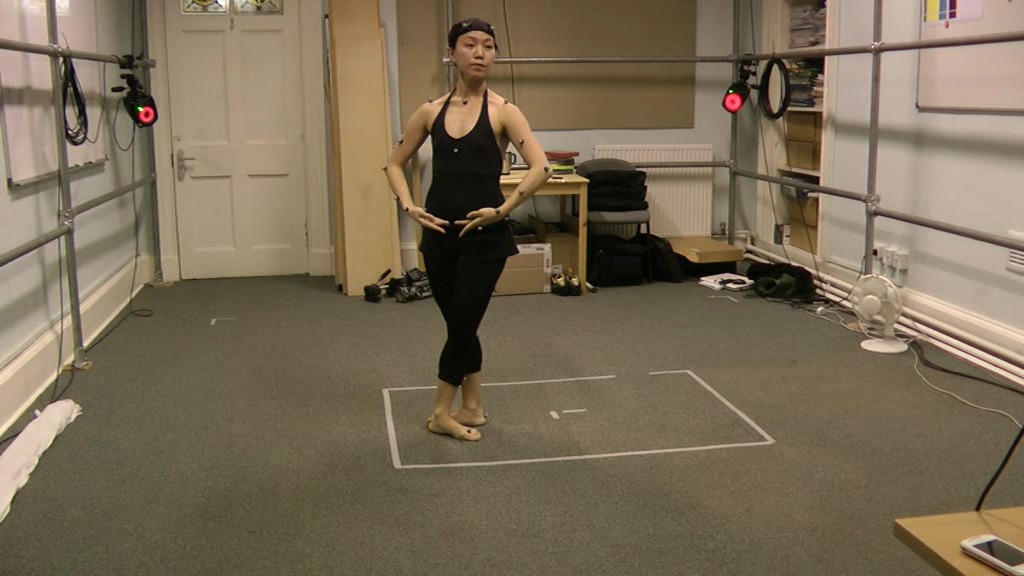}\hspace{7mm}
\includegraphics[height=1.9in]{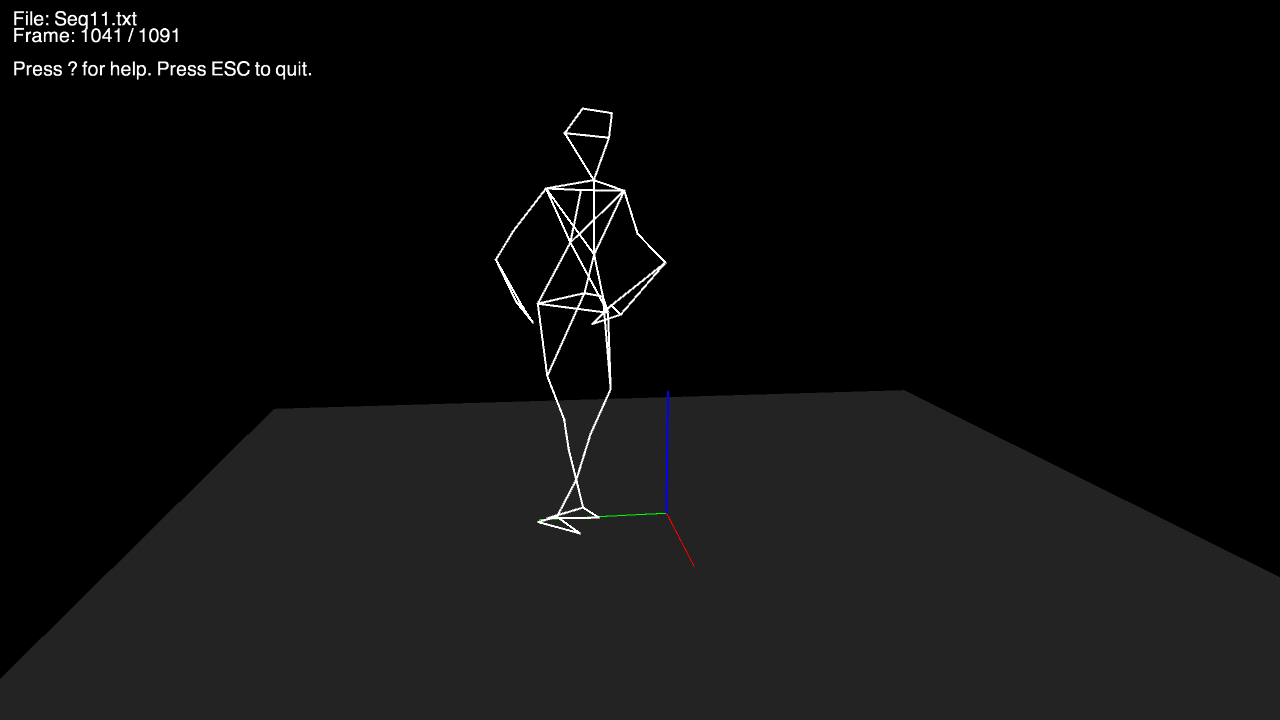}}
\end{picture}\\
\vspace*{-2mm}
\caption{Data collection procedure. 
 \textsc{Left:} An example of a motion capture situation on
  video. \textsc{Right:} A visualization of the captured pose.}
\label{fig:mocap1415}
\end{figure*}

\section{Data and Preprocessing}
\label{sec:data}

In order to study unconstrained performances without limiting ourselves to
specific tasks or parts of the body, we analyse motion capture data.
Motion capture data is typically
obtained by recording a subject by a set of cameras, and using
special-purpose image processing technologies to convert the recorded
video into variables such as 3D coordinates or angles of joints
(wrists, elbows, shoulders, waist, knees, etc).

\begin{table}[t!]
\caption{Summary of the data used in the experiments.
}
\label{tab:data}
\begin{tabular}{cp{67mm}r}
\# & \textsc{Label} & \textsc{$n$}\\
\hline
1 & \emph{adagio (temps li\'e, arabesque, pas de bourr\'ee, balanc\'e) } & 4254\\
2 & \multicolumn{1}{c}{\emph{---ii---}} & 4459\vspace{3pt}\\
3 & \emph{tomb\'e pas de bourr\'ee, Italian fouett\'e, piqu\'e turn, jet\'e en tournant } & 4001\\
4 & \multicolumn{1}{c}{\emph{---ii---}} & 3724\vspace{3pt}\\
5 & \emph{petit jet\'e (glissade jet\'e, ballott\'e, ballon, entrechat, assembl\'e) } & 1535\\
6 & \multicolumn{1}{c}{\emph{---ii---}} & 1574\vspace{3pt}\\
7 & \emph{grand jet\'e (battement d\'evelopp\'e, chass\'e, grande jet\'e d\'evelopp\'e, arabesque, 
fouett\'e saut\'e, jet\'e en tournant) } & 1560\vspace{3pt}\\
8 &  \multicolumn{1}{c}{\emph{---ii---}} & 1621\vspace{3pt}\\
9 & \emph{petit jet\'e (tendu crois\'e, sissonne devant ferm\'ee, derri\`ere ferm\'ee, sissonne ouv\'ert pas de bourr\'ee) } & 1091\\
10 & \multicolumn{1}{c}{\emph{---ii---}} & 1114\vspace{3pt}\\
\end{tabular}
\end{table}

For out experiment, we recorded the performance of a professional dancer
performing movement sequences of her own choice.
The recording and motion capture analysis was performed at the Perception, 
Action and Cognition Lab, University of Glasgow,
see Table~\ref{tab:data} and
Fig.~\ref{fig:mocap1415}. 
The sequences are recorded at
frame rate 120 per second. For each frame, the data contains $p=111$
features, corresponding to the three-dimensional coordinates of 37
markers attached to different parts of the body.

The inherent problem in predicting one motion sequence by another is
the possible misalignment of the sequences in
time.  Usually, even very carefully repeated movements are
slightly out of synchronization, and hence when predicting the
$t$'th frame of sequence $\mathbf{x}$, the most useful frame of
sequence $\mathbf{y}$ may not be the $t$'th frame but the
$t+\delta$'th one with $\delta \neq 0$. Therefore, it is necessary
to align the two sequences to obtain a better synchronization.

We aligned each pair of sequences in the data set by applying
Canonical Time Warping (CTW)\footnote{Matlab code is available at
  \url{www.humansensing.cs.cmu.edu/projects/ctwCode.html}.}~\cite{zho:tor:09},
a state-of-the-art technique for aligning sequences describing human
behavior.  CTW uses the more traditional Dynamic Time Warping
(DTW)~\cite{rab:jua:93} as an initial solution but improves it by
adopting features from Canonical Correlation Analysis (CCA)
(see~\cite{and:03}). This allows alignment based on a more flexible
concept of similarity than usually used in DTW. 

The result of a pairwise alignment of two sequences, with possibly
different lengths, is a new pair of aligned sequences whose lengths
are equal, such that each frame in one sequence matches as well as
possible with the same movement (similar measured features) in the
other. To achieve this, the CTW algorithm duplicates some of the
frames in each sequence so as to ``slow down'' the sequence in
question at suitable points; see the example in
Fig.~\ref{fig:mocap1415}. When measuring the throughput, we
skip the duplicated frames in sequence $\mathbf{x}$ in order to avoid
unnecessarily magnifying their impact.
Hence, if frame $t$ is duplicated in sequence
$\mathbf{x}$ so that in the aligned sequence, $\mathbf{x}'$, frames
$t$ and $t+1$ are identical, we skip the $t+1$'th frame (of both
$\mathbf{x}'$ and $\mathbf{y}'$) when evaluating the throughput,
Eq.~\eqref{eq:capa}. The sequences were also normalized
so that each feature has mean zero and unit variance.
It is important to also note that we compute the
residuals of both sequences from the unaligned sequences where there
are no duplicate frames. However, the alignment is done based on the
actual sequences (not the residuals).

As an undesirable consequence of the use of alignment methods in
preprocessing the motion capture data, we lose the information about
the temporal accuracy of the movements. 
Clearly, a significant amount of controlled information
are required for timing the motor responses. Working with aligned
sequences, there is no way to measure the accuracy to which the
repeated performance is synchronized with the original
performance. One possibility is to examine the alignment itself to see
how much information is required to bring the two sequences
in close agreement, and to add this information to the information
content due to spatial accuracy. We will explore this issue in
further work.

\begin{figure}[b!]
\includegraphics[width=\columnwidth]{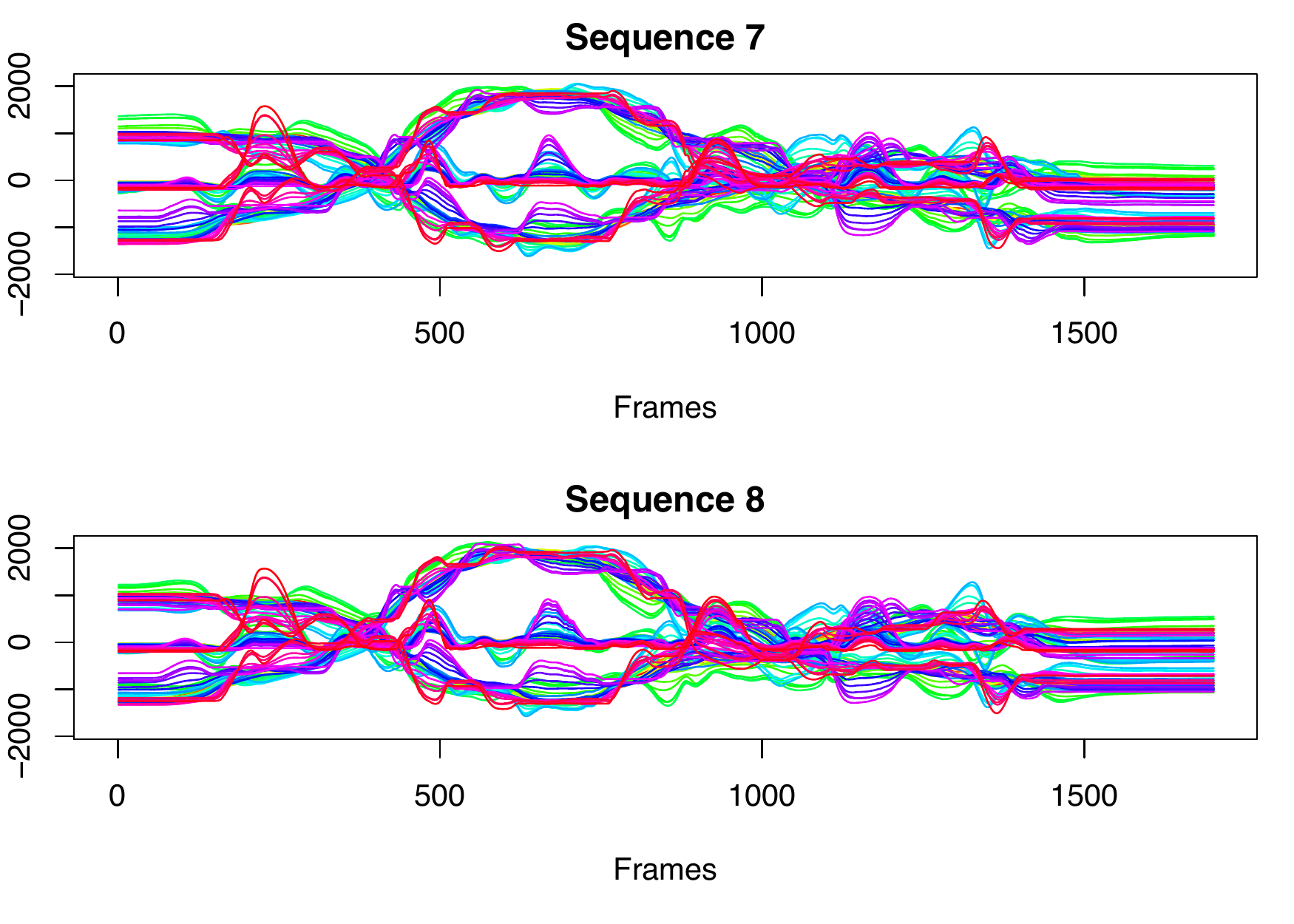}
\caption{
The plotted sequences of two motion capture
  sequences (sequences 7 and 8, see Table~\ref{tab:data}) after alignment --- note the high similarity of the two sequences.}
\end{figure}

\section{Results and Discussion}
\label{sec:results}

Table~\ref{tab:tp} lists all the throughput values for each pair of
movement sequences corresponding to the same movement pattern,
see Table~\ref{tab:data}.
Of all the pairwise throughput values, $\TP(\mathbf{x} \mid
\mathbf{y})$, the highest one, 1653 bits per second (bps), is obtained
for sequence 8 conditioned on sequence 7, see
Fig.~\ref{fig:mocap1415}. Their similarity is easily confirmed
visually from the video recordings and the animated reconstructions
available (not shown).  The values are nearly symmetric: the throughput
in sequence 7 conditioned on sequence 8 is 1580 bps. The lowest
throughput, 640 bps, was observed for sequence 1 conditioned on sequence 2.

\begin{table}[t!]
\caption{Measured throughput values for the sequences listed in
Table~\ref{tab:data}.
}
\label{tab:tp}
\center
\begin{tabular}{ccr}
\toprule
$\quad\mathbf{x}\quad$ & $\quad\mathbf{y}\quad$ & $\quad\TP(\mathbf{x}\mid\mathbf{y})\quad$\\
\midrule
1 & 2 & 640 bps \\
2 & 1 & 668 bps\vspace{3pt}\\ 
3 & 4 & 1408 bps \\ 	
4 & 3 & 1481 bps\vspace{3pt}\\
5 & 6  & 931 bps \\
6 & 5  & 914 bps\vspace{3pt}\\
7 & 8  & 1580 bps\\ 
8 & 7  & 1653 bps\vspace{3pt}\\ 
11 & 12 & 763 bps\\
12 & 11 & 756 bps \\
\bottomrule
\end{tabular}
\end{table}


As a sanity check, we also evaluated the throughput for pairs of
sequences that were not repetitions of the same movement pattern.
As expected, the obtained throughput values are all very small or
even negative\footnote{Negative
  values are possible due to the second term, ${1\over 2}\log_2 n$, in
  Eq.~\eqref{eq:muti}.  In terms of the Minimum Description Length
  (MDL) Principle~\cite{ris:78,gru:07}, this would be taken to
  indicate that a model where $\mathbf{x}$ and $\mathbf{y}$ are
  independent is superior to the model
  where they are correlated via the innovation sequences.
  Note that this is equivalent to
  model selection using the Bayesian Information Criterion
  (BIC)~\cite{sch:78}.}.




\section{Conclusions and Future Work}
\label{sec:conclusion}

The experiment we have described demonstrates the main idea in our
framework, i.e., extending the prevailing information-theoretic
framework to allow completely unconstrained movements, and thereby, to
determine the maximum of the achievable information capacity. Motion
capture data provides the best way to characterize such movements in a
way that does not rule out any potentially informative aspects in
them.


That said, it will be interesting to compare the capacity estimates
obtained by other methods, such as pointing devices (the traditional
tool in Fitts' paradigm), data gloves, etc., and to see if the earlier
results are replicated. For instance, it is interesting to see if more
information can be extracted from Fitts' original reciprocal pointing
task by recording the movements by a data glove or motion capture: the
question is whether the path along which the hand operating the
pointer moves between the two targets carries additional information
beyond the information provided by the end-points, and if it does, how
much.


Achieving the goal of constructing a complete and reliable
measure of information capacity will lead to a wealth of useful
knowledge about the human motor system. Concrete utility is to
be seen, for instance, in the study of novel human-computer interfaces
that involve free whole-body expression.  Possible applications in
sports science include training of complex motor schemas with
reference models.  Potential new diagnostic tools based on monitoring
changes in the information capacity of the motor system may offer
great societal value through early identification of
neurological disorders related to motor dysfunction and in monitoring
recovery of neuroplasticity after lesions.
We will explore these lines of research in further work.




\section*{Acknowledgments}

The authors Frank Pollick for the chance to use the motion capture 
system at the University of Glasgow, and Naree Kim for dancing.
We also thank Kristian Lukander for several discussions on the
measurement of information capacity, and Daniel Schmidt and Petri
Lievonen for useful comments on an earlier draft of the paper. Any
remaining errors are naturally due to the authors.  The research by TR, LL, and
AM was funded by the Academy of Finland under projects PRIME and MODEST,
and the Pascal Network-of-Excellence.  The research by AO was funded
by Emil Aaltonen Foundation, the Smart Spaces Thematic Action Line
of EIT ICT Labs, and the Max Planck Center for Visual Computing and
Communication (MPC-VCC).





\begin{thebibliography}{99}

\bibitem{acc:zha:99}
J.~Accott and S.~Zhai. ``Performance evaluation of input devices in 
trajectory-based tasks: an application of the steering law'',
\emph{Proc. CHI'07}, ACM Press, pp. 466--472, 1999.

\bibitem{and:03}
T.~W.~Anderson. \emph{An Introduction to Multivariate Statistical
Analysis}, Wiley, 2003.

\bibitem{atk:07}
P.~Atkinson. ``The best laid plans of mice and men: the computer
mouse in the history of computing'',
\emph{Design Issues} \textbf{23}:46--61, 2007.

\bibitem{cov:tho:06}
T.~Cover and J.~Thomas. \emph{Elements of Information Theory}, 2nd Ed.,
Wiley, 2006.

\bibitem{gru:07}
P.~Gr{\"u}nwald. \emph{The Minimum Description Length Principle},
MIT Press, 2007.

\bibitem{fit:54}
P.~M.~Fitts. ``The information capacity of the human motor system in
controlling the amplitude of movement'',
\emph{J Experim Psychology} \textbf{47}:381--391, 1954.

\bibitem{fit:pet:64}
P.~M.~Fitts and J.~R.~Peterson. ``Information capacity of discrete motor
responses'',
\emph{J Experim Psychology} \textbf{67}:103-–112, 1964.

\bibitem{jon:led:06}
L.~A.~Jones and S.~J.~Lederman. \emph{Human Hand Functioning},
Oxford University Press, 2006.

\bibitem{mac:89}
I.~S.~MacKenzie. ``A note on the information-theoretic basis for Fitts' law'',
\emph{J Motor Behavior} \textbf{21}:323-–330, 1989.

\bibitem{mac:92}
I.~S.~MacKenzie. ``Fitts' law as a research and design tool in 
human-computer interaction'',
\emph{Human-Computer Interaction} \textbf{7}:91-–139, 1992.

\bibitem{rab:jua:93}
L.~Rabiner and B.-H.~Juang. \emph{Fundamentals of Speech Recognition},
Prentice Hall, 1993.

\bibitem{rob:kav:76}
G.~H.~Robinson and R.~C.~Kavinsky. ``On Fitts' law with
two-handed movement'',
\emph{IEEE Trans Syst, Man \& Cybern}, \textbf{6}:504--505, 1976.

\bibitem{ris:78}
J.~Rissanen. ``Modeling by shortest data description'',
\emph{Automatica} \textbf{14}:445--471, 1978.

\bibitem{sch:78}
G.~Schwarz. ``Estimating the dimension of a model,''
\emph{Annals of Statistics} \textbf{6}:461--464, 1978.

\bibitem{sou:mac:04}
R.~W.~Soukoreff and I.~S.~MacKenzie. ``Towards a standard for pointing device evaluation, perspectives on 27 years of Fitts' law research in HCI'',
\emph{Int J Human-Computer Studies} \textbf{61}:751--789, 2004.

\bibitem{wel:68}
A.~T.~Welford. \emph{Fundamentals of Skill}, Methuen, 1968.

\bibitem{zha:02}
S.~Zhai. ``On the validity of throughput as a characteristic of 
computer input,'' 
\emph{IBM Research Report RJ 10253}, IBM Research Center, Almaden,
California, 2002.

\bibitem{zho:tor:09}
F.~Zhou and F.~de~la~Torre. ``Canonical time warping for alignment
of human behavior'', \emph{Advances in Neural Information Processing
Systems} (NIPS), 2009.

\end{thebibliography}
\end{document}